\documentclass[onecolumn, 12pt]{IEEEtran}

\usepackage{graphicx}

\begin{document}

\title{Preventing Time Synchronization in NTP's Broadcast Mode}

	

\author{Nikhil Tripathi\IEEEauthorrefmark{1},
        Neminath Hubballi\IEEEauthorrefmark{2}
       \thanks{Nikhil Tripathi is with Technical University of Darmstadt, Rheinstr. 75, 64295 Darmstadt, Germany. Neminath Hubballi is with Discipline of Computer Science and Engineering, Indian Institute of Technology Indore, India. (E-mails:\IEEEauthorrefmark{1}nikhil.tripathi@sit-extern.fraunhofer.de, \IEEEauthorrefmark{2}neminath@iiti.ac.in). \IEEEauthorrefmark{1}Corresponding Author}
}

\maketitle

\begin{abstract}
  Network Time Protocol (NTP) is  used by millions of hosts in Internet today to synchronize their clocks. Clock synchronization is necessary for many network applications to function correctly.  Unsynchronized clock may lead to failure of various core Internet services including DNS and RPKI based interdomain routing and opens path for more sophisticated attacks. In this paper, we describe a new attack which can prevent a client configured in  NTP's broadcast mode from synchronizing its clock with the server. We test the attack in real networks and show that it is effective in both authenticated and unauthenticated broadcast/multicast modes of NTP. We also perform experiments to measure the overall attack surface by scanning the entire IPv4 address space and show that NTP broadcast mode is being used in the wild by several low stratum (highly accurate) hosts. We also suggest few countermeasures to mitigate the proposed attack.
\end{abstract}




\section{Introduction}
\label{ntp_introduction}

Network Time Protocol (NTP) \cite{rfc5905} is one of the oldest protocols and is used to synchronize clocks of computer systems on the Internet. Clock synchronization among computers is important for many reasons as inconsistencies in clock time can affect various core Internet services such as Domain Name System (DNS) resolution, interdomain routing using Resource Public Key Infrastructure (RPKI) and Transport Layer Security (TLS) \cite{firetalk}, \cite{Mills:2010:CNT:1951866}, \cite{defcon}, \cite{malhotra_ndss}. Thus, if an adversary is able to manipulate clocks of important systems on the Internet by obstructing NTP's operation, services to a large number of the Internet users will be affected. There have been several incidences in the past where clock synchronization failure led to breakdown of various services on the faulty systems, all at the same time. For example, two important NTP servers of U.S. Naval Observatory (USNO) were sent back in time by about 12 years on November 19, 2012 \cite{ntp_nanog}, ``causing outages at a variety of devices including Active Directory (AD) authentication servers and routers" \cite{ntp_routers_microsoft}. NTP's potential for Distributed DoS (DDoS) using reflection and amplification techniques is well studied by the security community \cite{czyz2014taming}. However, like in other application layer protocols \cite{tripathi2015exploiting, tripathi2016secure, tripathi2018slow, hubballi2017closer}, researchers have now started paying attention to find vulnerabilities in NTP specification and implementation also \cite{malhotra}, \cite{malhotra_ndss}, \cite{aanchal}. 

In this paper, we describe a new attack that exploits a vulnerability present in NTP's broadcast mode. The proposed attack prevents a NTP client from synchronizing its clock with a NTP broadcast server by sending spoofed NTP packets to the NTP client and the broadcast server. We also show how the proposed attack is effective in both authenticated and unauthenticated broadcast modes of NTP. To the best of our knowledge, this is the first attack proposed against NTP protocol since launch of NTP's reference implementation's most recent version \textit{ntpd} v4.2.8p13 \cite{ntpd} on March 7th, 2019. As a first line of defense, we also suggest few countermeasures. In summary our contributions in this paper are:\\\\
\textbf{1.} We propose a new attack that can prevent a NTP client from synchronizing its clock with a NTP server. \\
\textbf{2.} We test the proposed attack in a real network and show that it is effective in both authenticated and unauthenticated NTP broadcast/multicast modes.\\
\textbf{3.} We perform extensive experiments to measure the attack surface on Internet by scanning the entire IPv4 address space and furnish the results. We also quantify the resulting attack surface and discuss that it is only the floor value of the actual attack surface on the Internet. \\
\textbf{4.} We suggest few countermeasures that can be deployed to effectively counter the proposed attack until the release of a proper security patch.
\\\\
\textbf{Attack Disclosure:} We have disclosed the attack to \textit{ntpd} developers on 2nd October, 2018. Also, CVE-2018-8956 has been reserved for the proposed attack.
\\
\textbf{Organization:} Rest of the paper is organized as follows. We give an overview of NTP protocol and its operation in Section \ref{ntp_background}. In Section \ref{ntp_attack}, we describe the working of proposed attack. In Section \ref{ntp_experiments}, we present the experiments performed to test the proposed attack against NTP's broadcast mode and the results of attack surface measurement in Internet. We describe some of the possible countermeasures to mitigate the proposed attack in Section \ref{countermeasures}. We explain previously known attacks against NTP and techniques to counter these attacks in Section \ref{ntp_literature}. Finally, the paper is concluded in Section \ref{ntp_conclusion}.

\section{Threat Model}
\label{threat_model}

In this section, we describe the threat model we consider in our work.

\section{Background}
\label{ntp_background}

NTP is in operation since 1985 and thus, is one of the oldest Internet protocols currently being used. Earlier versions of the protocol, NTPv1, NTPv2 and NTPv3 were described in RFC 1059 \cite{rfc1059}, RFC 1119 \cite{rfc1119} and RFC 1305 \cite{rfc1305} respectively while the current version of the protocol, NTPv4, is described in RFC 5905 \cite{rfc5905} which is an \textit{Internet Standards Track} document since June 2010. Despite of changing the protocol specification over the years to make the protocol more robust and accurate, various NTP implementations today do not completely follow the specification \cite{aanchal}. Instead, it is \textit{ntpd}, the reference implementation of the protocol that ``practically defines it" \cite{malhotra_ndss}. \textit{ntpd} has been updated/modified several times in the last few decades to make it secure against various attack vectors found in the implementation over time. \textit{ntpd} v4.2.8p13 is the most recent version launched on March 7th, 2019. 

\subsection{NTP Packet Structure}
\label{ntp_packet}

\begin{figure}[h]
	\begin{center}
		\scalebox{0.38}{\includegraphics{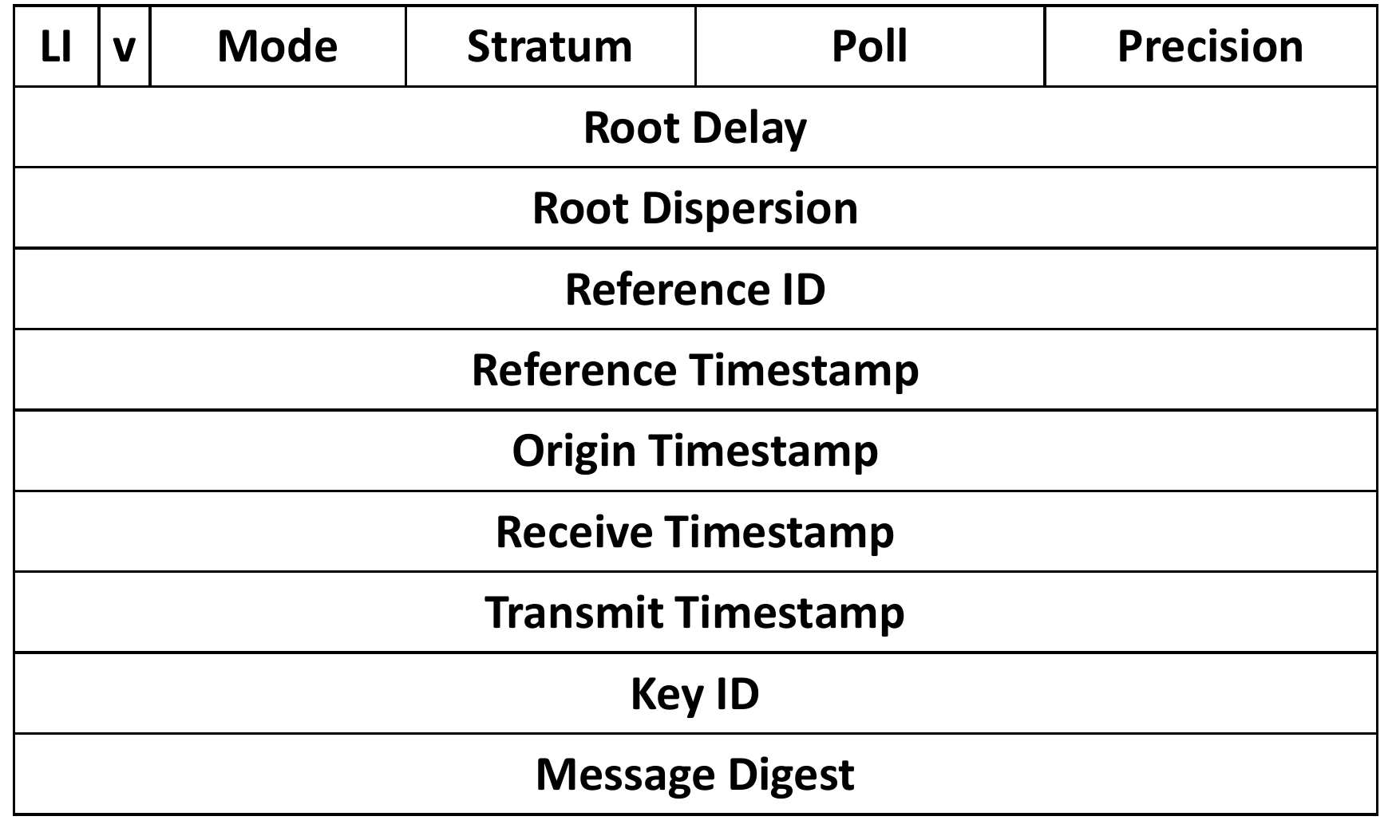}}
		\caption{NTP packet structure}
		\label{ntp_header}
	\end{center}
\end{figure}

Different types of NTP packets exchanged during clock synchronization process have a common packet structure. This structure is shown in Figure \ref{ntp_header}. 
In this subsection, we describe the fields of this packet structure essential to understand the working of our proposed attack. For a detailed explanation of NTP's packet structure, reader is referred to RFC 5905 \cite{rfc5905}.\\
\textbf{Mode:} This is a 3-bit field that represents the NTP mode (client/server, broadcast, etc.) to which the packet belongs.\\
\textbf{Stratum:} This is an 8-bit field that represents the stratum of the NTP system. Stratum is a simple measure of the synchronization distance from the primary time source and its value ranges from 0 to 16 inclusive.\\
\textbf{Poll:} This is an 8-bit field that represents the maximum interval between successive messages in log base 2 seconds. In case of a KoD packet (discussed later in Section \ref{kod}), this field represents the minimum amount of time for which an NTP client should refrain itself from querying the server.\\
\textbf{Reference ID:} This is a 32-bit field identifying a particular server or reference clock to which the NTP system is synchronized.\\
\textbf{Reference Timestamp:} This is a 32-bit field that represents the time when system clock was last updated.\\
\textbf{Origin Timestamp ($T_1$):} This is a 32-bit field that represents the time at the client when a NTP request departed for the server. In broadcast mode, since the client does not send any NTP request and it is the server that sends unsolicited mode 5 packets at regular intervals, this field is set to NULL.\\
\textbf{Receive Timestamp ($T_2$):} This is a 32-bit field that represents the time at the server when the request arrived from the client. Since there is no corresponding request sent by the client in broadcast mode, this field is also set to NULL.\\
\textbf{Transmit Timestamp ($T_3$):} This is a 32-bit field that represents the time at the server when the NTP packet left for the client. \\
\textbf{Key ID:} This is a 32-bit field that identifies the key used to authenticate the packet. There are multiple keys that can be configured on a NTP device and each key is identified by the Key ID. \\
\textbf{Message Digest:} This is a 128-bit field that contains the MD5 hash calculated over the contents of the NTP packet using a symmetric key $k$.

\subsection{NTP Operation}
\label{ntp_working}

\begin{figure}[h]
	\begin{center}
		\scalebox{0.4}{\includegraphics{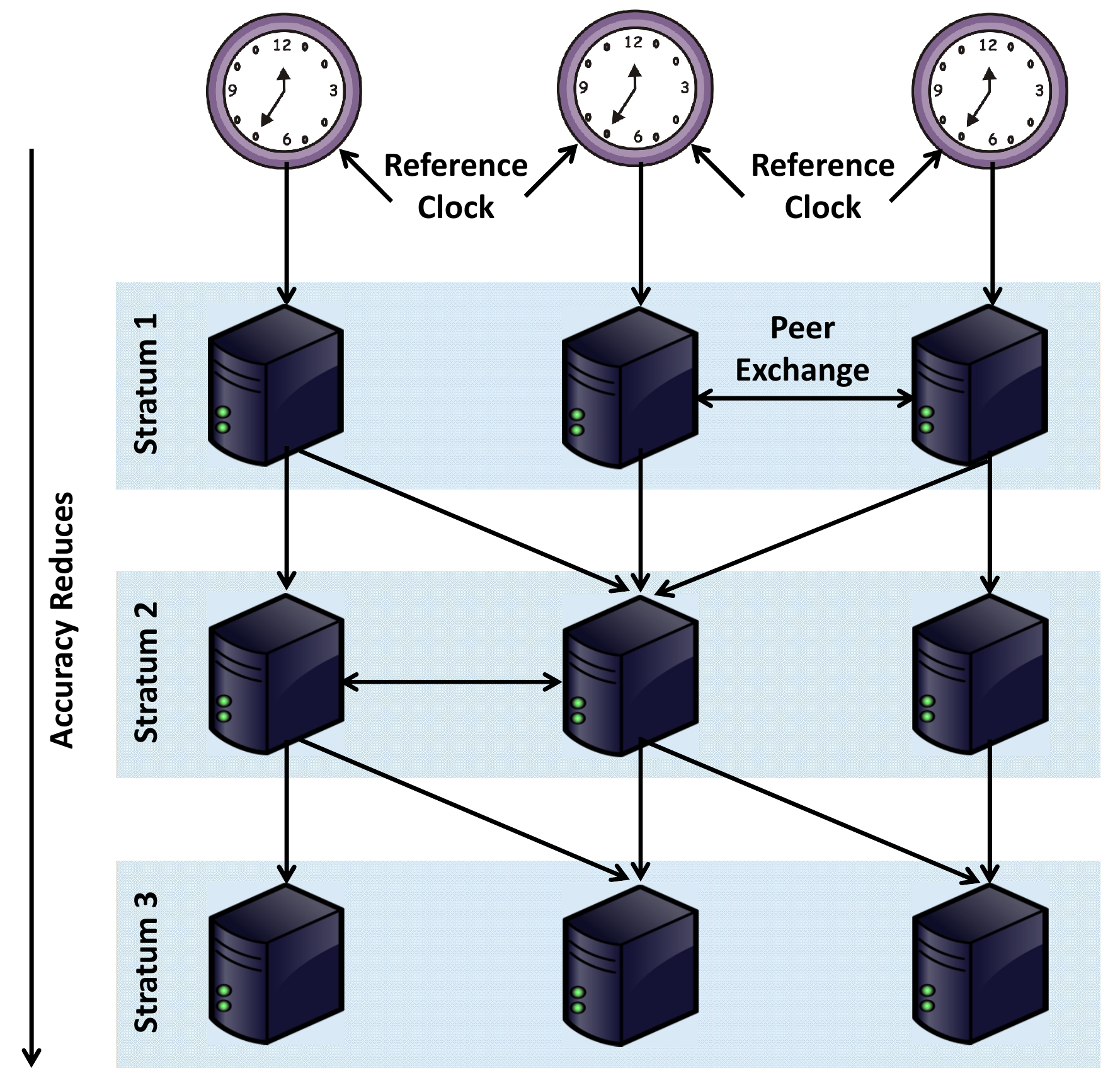}}
		\caption{NTP hierarchy}
		\label{ntp_hierarchy}
	\end{center}
\end{figure}

NTP is designed to operate under several modes: \textit{client/server} (mode 3/mode 4), \textit{broadcast} (mode 5), \textit{multicast} (mode 5) and \textit{symmetric} (mode 1 and mode 2) modes. All these modes are recommended to be processed by the same codepath of the NTP's implementation \cite{rfc5905}.
NTP's most common mode of operation is client/server mode in which an NTP client sends timing queries to a set of static NTP servers which are configured manually. Stratum $s$ NTP systems act as servers for stratum $s+1$ systems and provide time information to them. There are 16 levels in this stratum hierarchy such that stratum 0 refers to reference clocks which receive time from satellite navigation system while stratum 1 refers to those NTP systems which are at the root of the NTP hierarchy and get the time directly from reference clocks and thus, are the most accurate NTP systems. Also, a failure to synchronize clock is represented as stratum 16. Thus, as we go down the hierarchy, the time accuracy reduces as shown in Figure \ref{ntp_hierarchy}. 
Also, lower stratum NTP servers are given preference and large number of NTP clients usually synchronize their clock with these servers. In NTP's broadcast mode, a set of clients listen to a broadcast server that periodically broadcasts timing information while in symmetric mode, NTP peers exchange timing information with each other. 

\subsection{NTP's Broadcast Mode Operation}
\label{ntp_broadcast}

NTP's broadcast mode is mostly used in environments having few NTP servers and large number of potential clients. An NTP server can be configured to periodically broadcast what is known as \textit{mode 5} NTP packets in the network. Similarly, NTP clients can be configured to accept and process these mode 5 packets. The broadcast server sends persistent broadcast association packets to the client and never demobilizes the association. On receiving these packets, the client creates an ephemeral association with the server that exists unless there is an error or timeout. While mobilizing the association, the broadcast NTP client changes its mode to client/server mode and calculates the \textit{path propagation delay} between client and server. On page 6 of RFC 5905 \cite{rfc5905}, it is mentioned that: 

\begin{quote}
	\textit{``It is useful to provide an initial volley where the client operating in client mode exchanges several packets with the server, so as to calibrate the propagation delay..."}
\end{quote}

While calculating path propagation delay in client/server mode, the client sends a \textit{mode 3} NTP query to the broadcast server. This query is sent by a client to obtain clock information from a server in order to synchronize its clock with the server's clock. On receiving the mode 3 query, the server sends \textit{mode 4} response to the client. This response is sent by the server to provide its clock information to the client. Using this response, the client calculates the path propagation delay and reverts back to broadcast mode and creates an association with the broadcast server upon receiving subsequent mode 5 packets. After this, the client regularly synchronizes its clock with the help of mode 5 packets periodically sent by the broadcast server.

\subsection{Built-in Security Mechanisms in NTP}
\label{ntp_builtin}

NTP facilitates some built-in security mechanisms to counter variety of attacks against the protocol. These are as follows:

%

\subsubsection{TEST2 \cite{rfc1305}}

NTP protocol requires the clients to check that the origin timestamp ($T_1$) present in a server's mode 4 response is the same as the transmit timestamp ($T_3$) present in the corresponding mode 3 query sent by the client earlier. In this way, an off-path malicious client, that cannot observe the communication between the NTP client and the server, does not know the timestamp and thus can not spoof the packets. This is called as TEST2. It is to be noted that all NTP packets have UDP source port 123 instead of random source port numbers. That is the reason port number can not be used as a nonce to match a mode 4 response with the corresponding mode 3 query.

\subsubsection{Panic Threshold and Step Threshold \cite{rfc5905}}

RFC 5905 defines a \textit{panic threshold} PANICT and suggests that if the time difference (known as \textit{offset}  and discussed later in Section \ref{ntp_vulnerability}) between an NTP client's clock and an NTP server's clock to which the client is attempting to synchronize is greater than PANICT, NTP program should quit. On page 44 of RFC 5905, it is mentioned that:
\begin{quote}
	\textit{``PANIC means the offset is greater than the panic threshold PANICT (1000 s) and SHOULD cause the program to exit"}
\end{quote}
This threshold prevents such attacks where a malicious client can attempt to tell a client to alter its local clock by a very large value so that the malicious client can create scenarios such as DNS cache and TLS certificate validity expiry which require time to be shifted by big steps (days or even years) in the past or future. RFC 5905 also defines a \textit{step threshold} and suggests that a NTP client should accept a time shift larger than the step threshold (125 ms by default) but smaller than PANICT (1000 s by default). 

\subsubsection{Kiss-o'-Death (KoD) Packet \cite{rfc5905}}
\label{kod}

RFC 5905 describes the use of a special packet called KoD to inform clients to stop sending packets that violate server access controls. For this purpose, the protocol specification defines several kiss codes \cite{rfc5905} which must be inspected by recipient clients and take the action accordingly. One such kiss code is 'RATE' which asks a recipient client to reduce the rate with which it is sending mode 3 queries to the server. On page 24 of RFC 5905, it is mentioned that: 
\begin{quote}
	\textit{``Rate exceeded. The server has temporarily denied access because the client exceeded the rate threshold."}
\end{quote}

\subsubsection{Authenticated NTP's Broadcast Mode}

An NTP client listens to broadcast mode 5 packets sent by any NTP broadcast server in the network so an adversary can send mode 5 NTP packets from a spoofed IP address and the client accepts it. Moreover, the origin timestamp field in mode 5 NTP packet sent by broadcast server is always NULL as NTP client does not send any query for broadcast packets. As a result, TEST2 that checks if the origin timestamp and the transmit timestamp present in the most recent query client sent to the server is same, is not applicable in the broadcast mode. Due to this, NTP client is unable to validate the received mode 5 packets. Thus, NTP's broadcast mode is vulnerable to identity spoofing attacks where malicious client can spoof the IP address of a broadcast server and send packets on its behalf. To address this issue, RFC 5905 \cite{rfc5905} recommends that NTP implementations should have some authentication mechanism to validate mode 5 NTP packets. On page 32 of RFC 5905, it is mentioned that:

\begin{quote}
	\textit{``There is no specific requirement for authentication; however, if authentication is implemented, then the MD5-keyed hash algorithm described in [RFC1321] must be supported."}
\end{quote}

The authentication scheme appends an MD5 hash obtained by using a symmetric key $k$ and NTP packet contents to the Message Digest field of NTP packet as shown in Figure \ref{ntp_header}.

\section{Proposed Attack}
\label{ntp_attack}

We found a vulnerability (CVE-2018-8956) in NTP's broadcast mode which can be exploited to launch an attack that can prevent a client configured in broadcast mode from synchronizing its clock with an NTP broadcast server. In this section, we first discuss the protocol vulnerability and how it can be exploited to launch the attack. Subsequently, we describe the procedure of attack execution in NTP's unauthenticated and authenticated broadcast modes.


\subsection{Vulnerability in the Protocol}
\label{ntp_vulnerability}

NTP packets exchanged between a client and server include four timestamps - Origin Timestamp ($T_1$) that determines client's time when it sent the request to the server, Receive Timestamp ($T_2$) that determines server's time when it received the request sent by the client, Transmit Timestamp ($T_3$) that determines server's time when it sent the response to the client and Destination Timestamp ($T_4$) that determines client's time when it received the response sent by the server\footnote{$T_4$ is not present in the NTP packet structure and the client retrieves this timestamp directly from its own clock.}. NTP uses these timestamps to calculate \textit{Offset} ($\theta$) which represents the time difference between the client's and server's clock and is given by 
\begin{equation}
\label{offset}
\theta=\frac{1}{2}((T_2-T_1)+(T_3-T_4))
\end{equation}
If the calculated offset $\theta$ is greater than the panic threshold PANICT, according to RFC 5905, \textit{ntpd} should quit with a diagnostic message to the system log. However, this creates surface for on-path \textit{Small-step-big-step attack} \cite{malhotra_ndss}. This issue was disclosed in CVE-2015-5300 and \textit{ntpd} v4.2.8p5 and versions released later were patched to prevent this attack. We also tested four most recent versions of \textit{ntpd} - v4.2.8p10, v4.2.8p11, v4.2.8p12 and v4.2.8p13 - and found that these versions do not quit on receiving a broadcast mode 5 packet whose $T_3$ is set to a timestamp such that the calculated $\theta>$ PANICT. However, the change in the implementation creates surface for another vulnerability as these versions start recalculating path propagation delay by exchanging mode 3 and mode 4 packets with the NTP server on receiving mode 5 packets whose $T_3$ is set to a timestamp such that the calculated $\theta>$ PANICT. Thus, to exploit this vulnerability, a malicious client can send such mode 5 packets to the victim client by spoofing the IP address of the genuine broadcast server. On receiving these mode 5 packets, the victim client starts sending mode 3 packets to recalculate the path propagation delay while on the other hand, the malicious client keeps sending large number of mode 3 packets to the broadcast server by spoofing the IP address of the victim client. As a result, the broadcast server starts responding with KoD packets (instead of mode 4 responses) having 'RATE' code in response to the genuine mode 3 packets sent by the broadcast client. Since broadcast client does not receive mode 4 responses from the server, it is not able to calculate the path propagation delay. As a result, the broadcast client is not able to synchronize its clock with the broadcast server.

Here arises a question - Can malicious client simply send spoofed KoD packets to the victim client as if they are from the broadcast server instead of inducing them from broadcast server to the victim client? The answer to this is \textit{ntpd} v4.2.8p4 and later implementations use origin timestamp ($T_1$) as a nonce to authenticate KoD packets \cite{malhotra_ndss}. Since malicious client (assuming \textit{off-path}) does not know this nonce, it can not send KoD packets to the victim client which are spoofed to look like they are coming from the broadcast server. In fact this was a vulnerability in \textit{ntpd} v4.2.8p3 and prior implementations which accept a KoD packet even if its origin timestamp is bogus \cite{malhotra_ndss}. After disclosure, the vulnerability was patched in \textit{ntpd} v4.2.8p4 and later implementations.

\subsection{Preventing Time Synchronization in Unauthenticated Broadcast Mode}
\label{unauth}

\begin{figure}[h]
	\begin{center}
		\scalebox{0.33}{\includegraphics{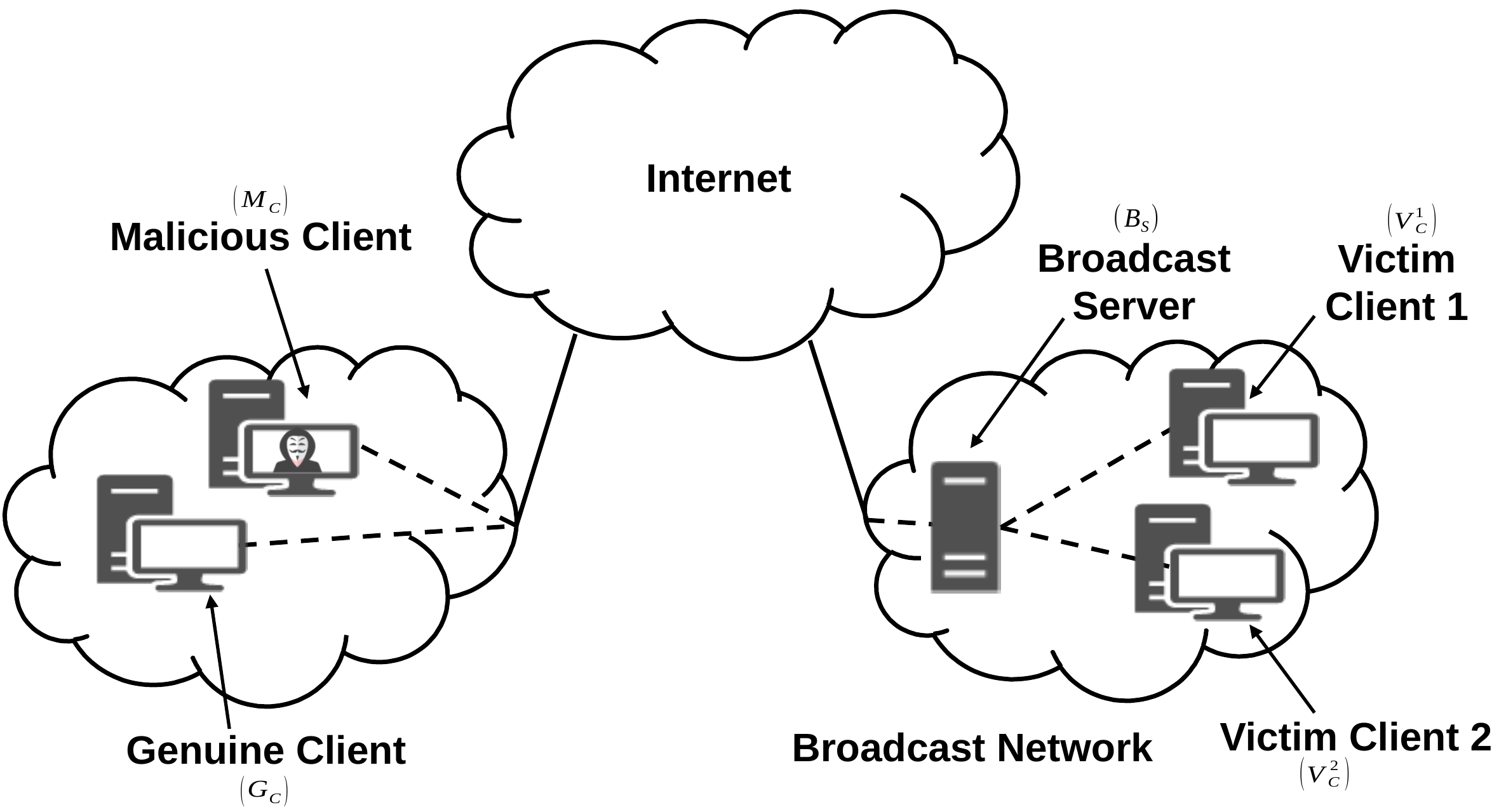}}
		\caption{Topology for attack description}
		\label{topology}
	\end{center}
\end{figure}

Consider a network topology having five entities as shown in Figure \ref{topology} - one $i^{th}$ stratum NTP broadcast server ($B_s$), two $(i+1)^{th}$ stratum victim clients ($V^1_c$ and $V^2_c$) configured to accept mode 5 NTP packets, one off-path malicious client ($M_c$) and one genuine client ($G_c$)- such that $B_s$, $V^1_c$ and $V^2_c$ are part of the broadcast network $N_1$ while $M_c$ and $G_c$ are part of another network $N_2$ on the Internet. Since the victim clients and malicious client are part of different networks in this scenario, the malicious client is \textit{\textbf{off-path}}.
Here, $G_c$ is controlled by malicious client and its only purpose is to confirm that $V^1_c$ is unable to synchronize its clock with $B_s$ while the attack is going on.
\\\\
\textbf{Launching the Attack:} The steps involved while launching the attack are as follows:
\\\\
\textbf{1.} In the first step, $V^1_c$ is let to successfully synchronize its clock with $B_s$. $V^1_c$ continuously listens to the mode 5 packets sent by $B_s$ at regular intervals. As soon as it receives the first mode 5 packet, it calculates the path propagation delay by exchanging mode 3 and mode 4 packets with $B_s$. After calculating the path propagation delay, $V^1_c$ reverts back to the broadcast mode and synchronizes its clock successfully.
\\
\textbf{2.} $M_c$ sends mode 5 packets to $V^1_c$ at regular intervals which are spoofed to look like they are coming from $B_s$. $T_1$ and $T_2$ in these mode 5 packets are set to zero while $T_3$ is set to a timestamp such that $\theta$ calculated by client is greater than PANICT. Few moments later, $M_c$ starts sending mode 3 packets to $B_s$ at regular intervals which are spoofed to look like they are coming from $V^1_c$.
\\
\textbf{3.} On receiving spoofed mode 5 packets sent by $M_c$ in previous step, NTP daemon of $V^1_c$ calculates $\theta$ using Equation \ref{offset}. Since $\theta>$ PANICT, the daemon starts sending mode 3 NTP queries to $B_s$ to recalculate the path propagation delay.
\\
\textbf{4.} Since $B_s$ continuously receives spoofed mode 3 NTP queries sent by $M_c$ in Step 2, $B_s$ sends KoD packets instead of valid mode 4 packets to $V^1_c$ in response to its mode 3 queries sent in Step 3 due to which $V^1_c$ is not able to calculate the path propagation delay. As a result, $V^1_c$ is not able to synchronize itself with $B_s$. 

\begin{figure}[h]
	\begin{center}
		\scalebox{0.45}{\includegraphics{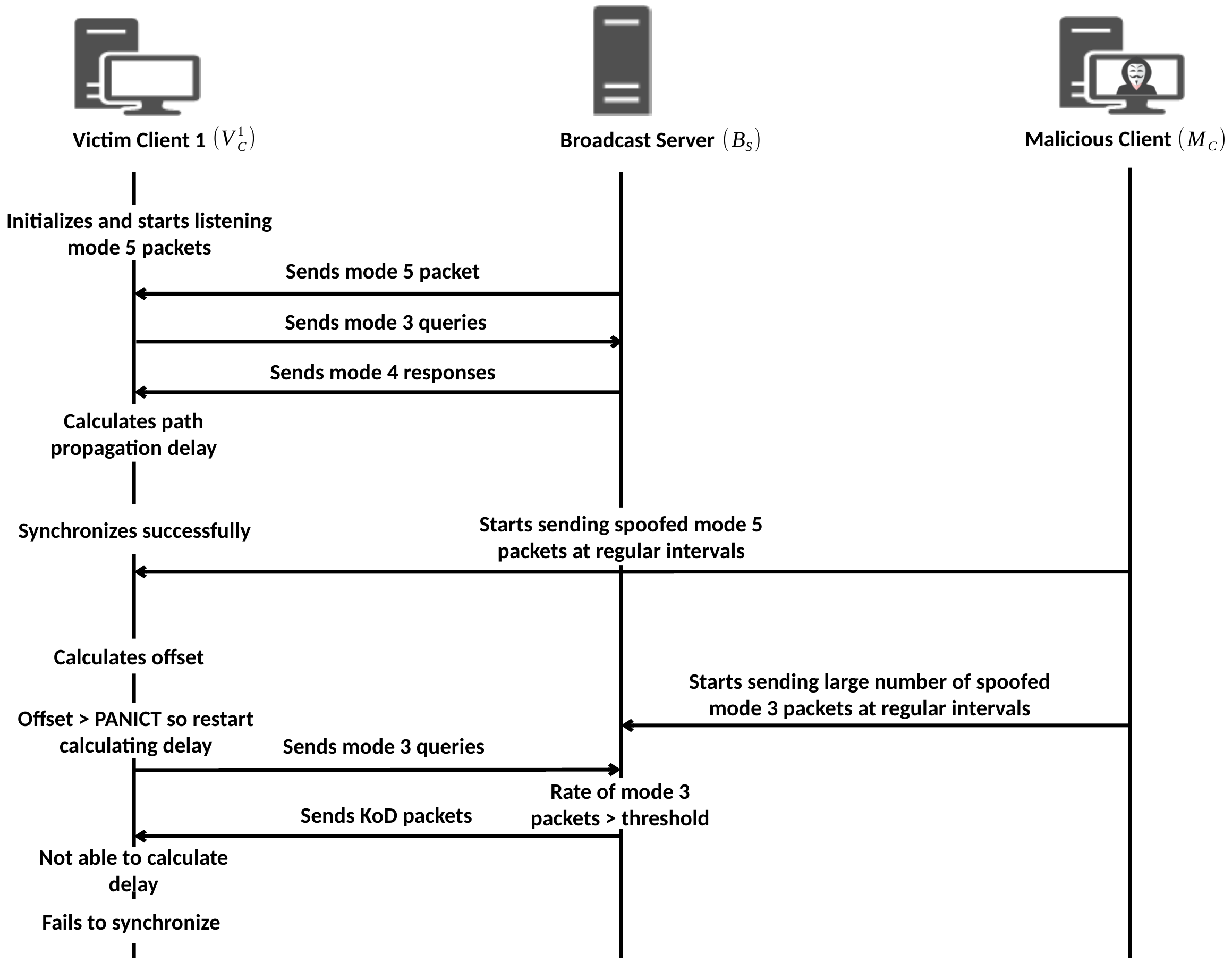}}
		\caption{Exchange of messages while launching the attack}
		\label{ntp_perspective_diagram}
	\end{center}
\end{figure}

The exchange of various messages and their respective time of transmission for the successful execution of proposed attack are shown in Figure \ref{ntp_perspective_diagram}.
\\\\
\textbf{Confirming Attack Success:} As per protocol specification \cite{rfc5905}, every NTP host acts as a client for a set of servers and as a server for a set of clients. For example, in Figure \ref{ntp_hierarchy}, an NTP host at stratum 2 acts as a client for a stratum 1 NTP host but as a server for a stratum 3 NTP host. Thus if an NTP host receives a mode 3 query from an NTP client attempting to synchronize its clock, it acts as a server and sends back a mode 4 response to the client. This mode 4 response contains a field \textit{reference timestamp} as shown in Figure \ref{ntp_header}. This field represents the time when system clock was last updated. In order to confirm that $V^1_c$ is no longer able to synchronize itself with $B_s$, $G_c$ sends genuine mode 3 NTP queries to $V^1_c$ at regular intervals and in return, $V^1_c$ sends back mode 4 NTP responses. In all these responses, if reference timestamp remains same, it means that $V^1_c$ is not able to synchronize itself with $B_s$.
\\\\
\textbf{Targeting other clients:} Following the above steps and by placing appropriate source and destination IP addresses in spoofed mode 3 and mode 5 packets, an adversary can attack any client in the target broadcast domain.


\subsection{Preventing Time Synchronization in Authenticated Broadcast Mode}
\label{auth}

\begin{figure}[h]
	\begin{center}
		\scalebox{0.33}{\includegraphics{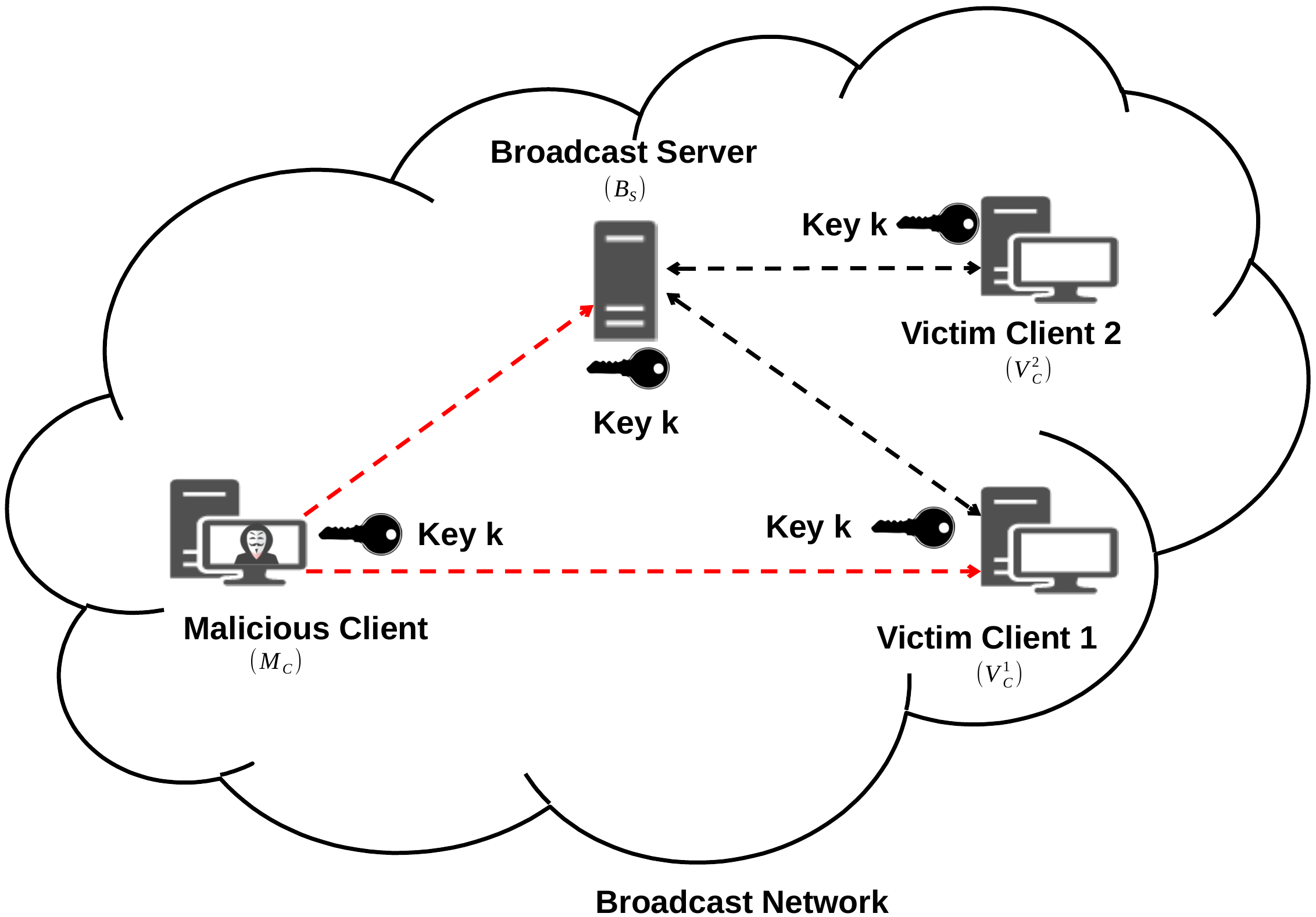}}
		\caption{Malicious client is part of the broadcast network}
		\label{ntp_all_part_of_same_network}
	\end{center}
\end{figure}

\begin{figure}[h]
	\begin{center}
		\scalebox{0.33}{\includegraphics{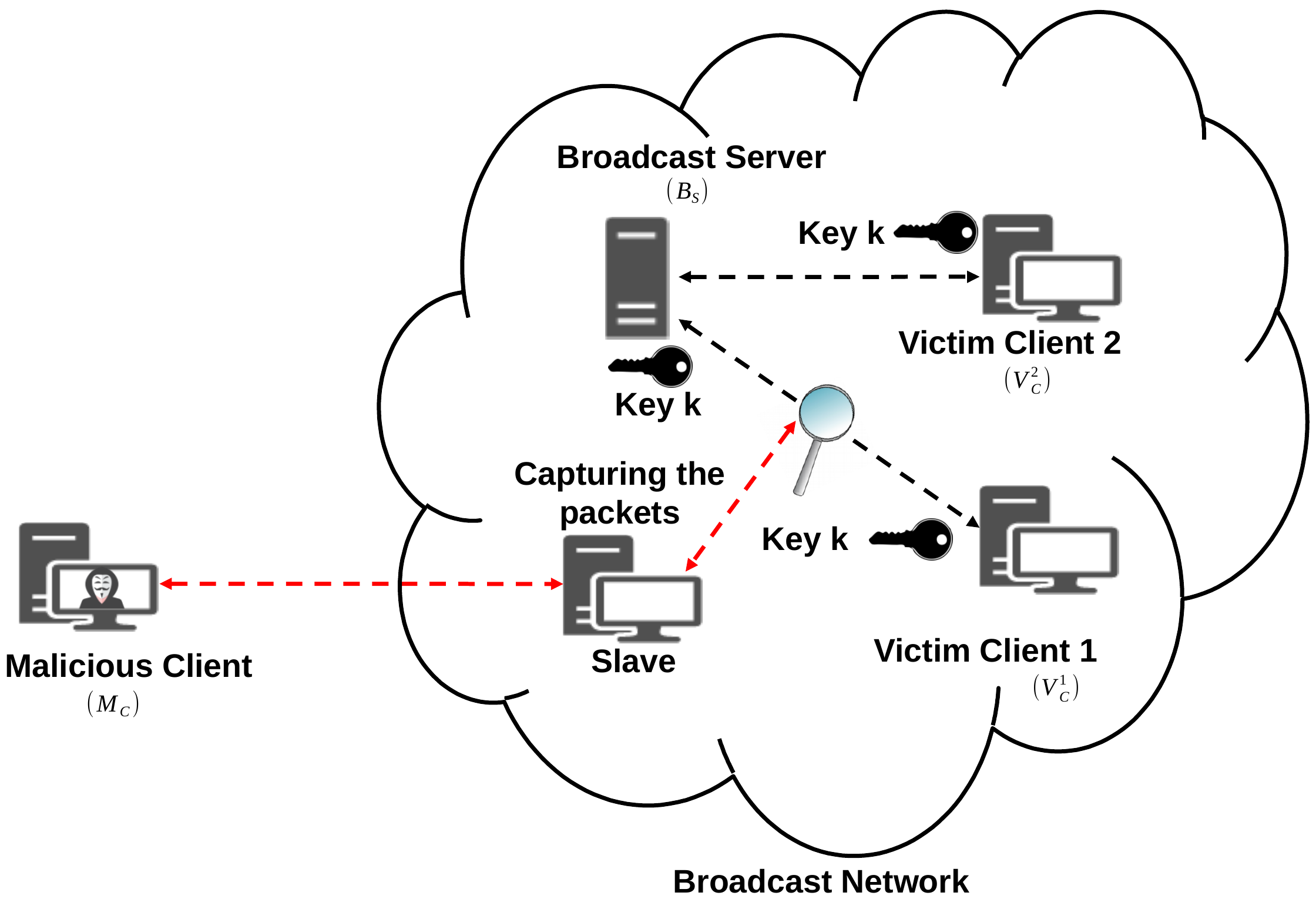}}
		\caption{Malicious client controls a slave in the broadcast network}
		\label{ntp_malicious_not_part_of_same_network}
	\end{center}
\end{figure}

\begin{figure}[h]
	\begin{center}
		\scalebox{0.45}{\includegraphics{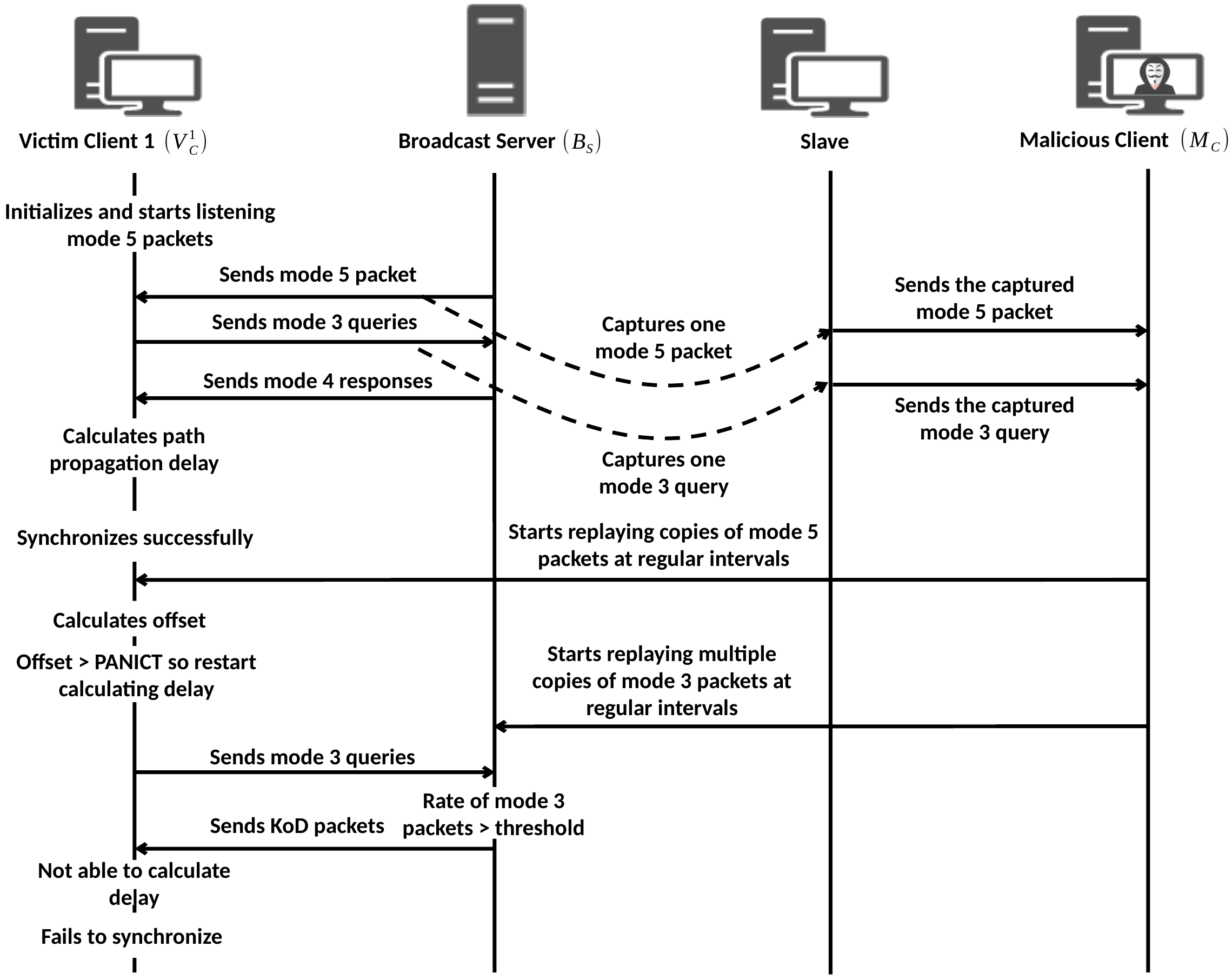}}
		\caption{Exchange of messages while launching the attack in second scenario}
		\label{ntp_malicious_not_part_timeline}
	\end{center}
\end{figure}

To describe the attack procedure in NTP's authenticated broadcast mode, we consider two different scenarios - 1) $M_c$ is part of the broadcast network (i.e. \textbf{\textit{on-path}}) $N_1$ as shown in Figure \ref{ntp_all_part_of_same_network} and 2) $M_c$ is not part of the broadcast network (i.e. \textbf{\textit{off-path}}) but it controls a slave $S$ in the broadcast network $N_1$ as shown in Figure \ref{ntp_malicious_not_part_of_same_network}. 
We now discuss how the proposed attack works in both these scenarios.
\\\\
\textbf{1. When $M_c$ is part of the broadcast network $N_1$:}
In this scenario, since $M_c$ is part of the broadcast network to which $B_s$, $V^1_c$ and $V^2_c$ belong, it also possesses the key used to authenticate NTP messages \cite{rfc7384}, \cite{draft}. Thus, $M_c$ can spoof identity of $B_s$ and send broadcast mode 5 packets containing bogus time to $V^1_c$ and $V^2_c$. Similarly, $M_c$ can also spoof identities of $V^1_c$ and $V^2_c$ to send large number of mode 3 packets to $B_s$. These packets are accepted by the destination entities as the MD5 hash appended to the packets are generated using $k$. Thus, by following the attack procedure discussed in Section \ref{unauth}, $M_c$ can prevent both $V^1_c$ and $V^2_c$ from synchronizing their clock with $B_s$.
\\\\
\textbf{2. When $M_c$ controls a slave in $N_1$:} In this scenario, $M_c$ controls a slave in $N_1$ that can capture one copy each of type mode 5 packet\footnote{Since mode 5 packet is a broadcast packet, the slave also receives this packet as it is part of the same network.} sent by $B_s$ and mode 3 query sent by either $V^1_c$ or $V^2_c$ to $B_s$\footnote{Unlike unauthenticated mode, the malicious client needs a copy of these packets to meet authentication check with $k$.}. Once slave captures the required packets, it forwards them to $M_c$. After receiving copies of mode 3 and mode 5 packets, $M_c$ can launch the attack from anywhere in the world but at a time $T_f$ such that $\theta$ calculated by $V^1_c$ at $T_f$ is greater than PANICT. To target $V^1_c$, $M_c$ continuously sends copies of mode 5 and mode 3 queries at regular intervals. The copies of mode 5 NTP packets are sent to IP address of $V^1_c$ with source IP address of $B_s$ while copies of mode 3 query are sent to $B_s$ with source IP address of $V^1_c$. Since $M_c$ does not change any field in the NTP header of captured mode 5 and mode 3 packets, the destination entities of these packets accept it as the MD5 hash appended to the packets is correct. Thus, the attack procedure discussed in Section \ref{unauth} prevents $V^1_c$ from synchronizing itself with $B_s$ in NTP's authenticated broadcast mode also. Likewise, $M_c$ can prevent $V^2_c$ from synchronizing itself with $B_s$ by placing appropriate source and destination IP address in IP header of mode 5 and mode 3 NTP packets. The exchange of various messages and their respective time of transmission for the successful execution of proposed attack against $V^1_c$ are shown in Figure  \ref{ntp_malicious_not_part_timeline}.


\section{Experiments and Results}
\label{ntp_experiments}

We tested the proposed attack in a real network setup. In this section, we first describe the experiments performed to test the proposed attack and then we present the attack surface results obtained from scanning the entire IPv4 address space on Internet.

\subsection{Testbed Setup}

\begin{figure}[h]
	\begin{center}
		\scalebox{0.33}{\includegraphics{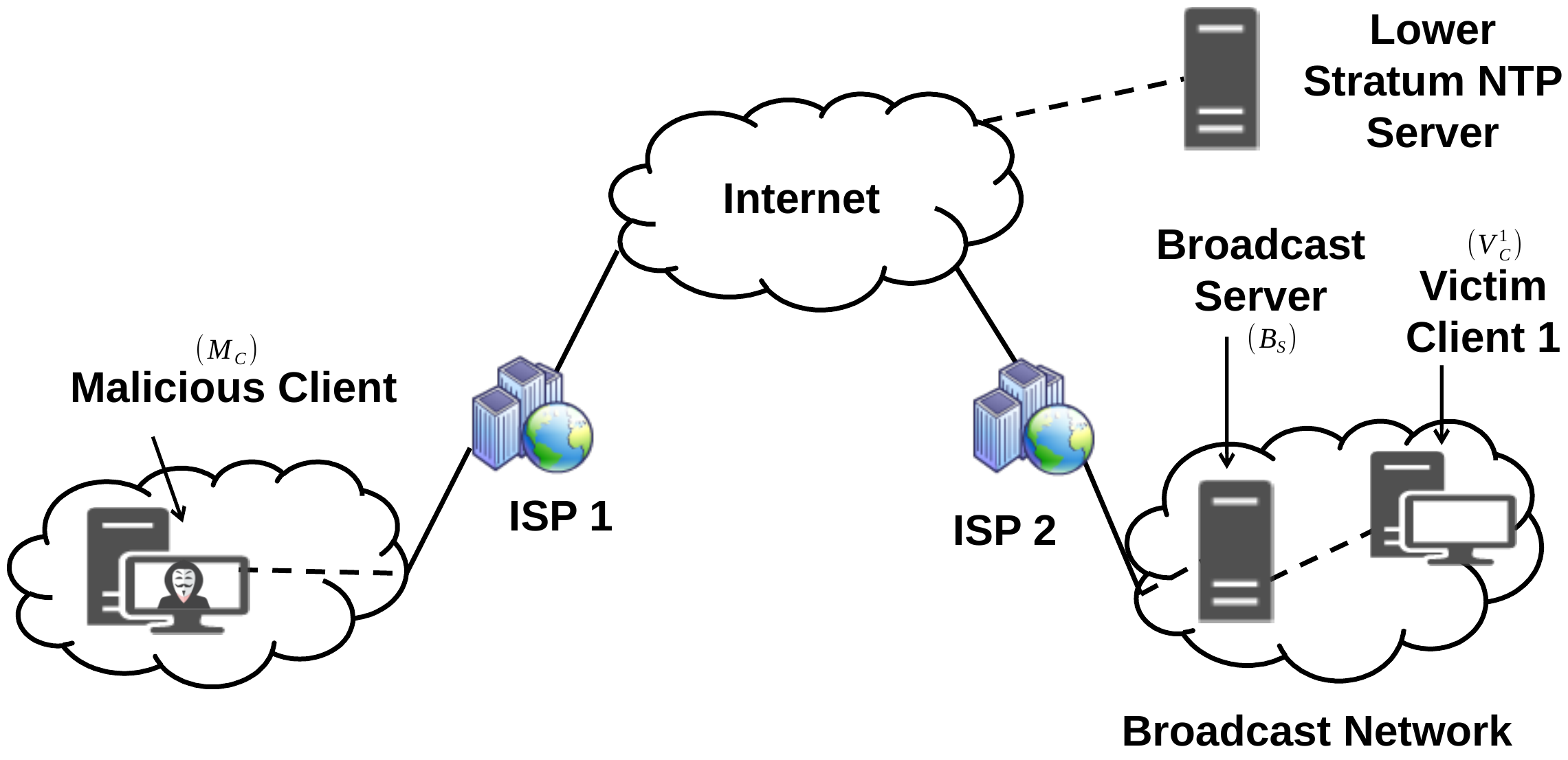}}
		\caption{Testbed setup}
		\label{ntp_attack_testbed_architecture}
	\end{center}
\end{figure}

The setup on which we tested our proposed attack is shown in Figure \ref{ntp_attack_testbed_architecture}.
It had three entities - an off-path malicious client, victim client and broadcast server. The broadcast server and victim client were part of same broadcast network while the malicious client was connected to another network. All the entities in the testbed were configured on computers running Kali Linux 2.0 operating system and having Core i7 processor with 4 GB of physical memory. The details about various entities and how they were configured are as follows:\\\\
\textbf{1. Broadcast Server:} We installed \textit{ntpd} v4.2.8p13 on the broadcast server and configured it by adding \texttt{broadcast} keyword and network broadcast address in NTP configuration file. A symmetric key was also configured in the configuration file for authentication purpose. This server was synchronized to an external stratum 2 NTP server on the Internet using one of its network interface while another interface was used to send mode 5 NTP packets on the internal network.\\\\
\textbf{2. Victim Client:} We also installed \textit{ntpd} v4.2.8p13 on the victim client and configured it by adding \texttt{broadcastclient} keyword and the required symmetric key in its NTP configuration file.\\\\
\textbf{3. Malicious Client:} The malicious client had two python scripts to send mode 5 and mode 3 NTP packets at regular intervals. In particular, we used \textit{Scapy} \cite{scapy_documentation} to send these packets. The first python script was configured to send copies of  mode 5 packet at a rate of 1 packet per second while the second python script was configured to first send two mode 3 packets to broadcast server and then at a rate of only 1 packet per 10 seconds.

Before the attack execution started, we let the victim client to synchronize its clock with the broadcast server. In the meantime, we captured one packet each of type mode 5 and mode 3 from the communication between broadcast server and victim client. These packets were then provided to the designated malicious client so that the packets could be sent by it later. 

\subsection{Attack Execution} 

\begin{figure*}
	\begin{center}
		\scalebox{0.53}{\includegraphics{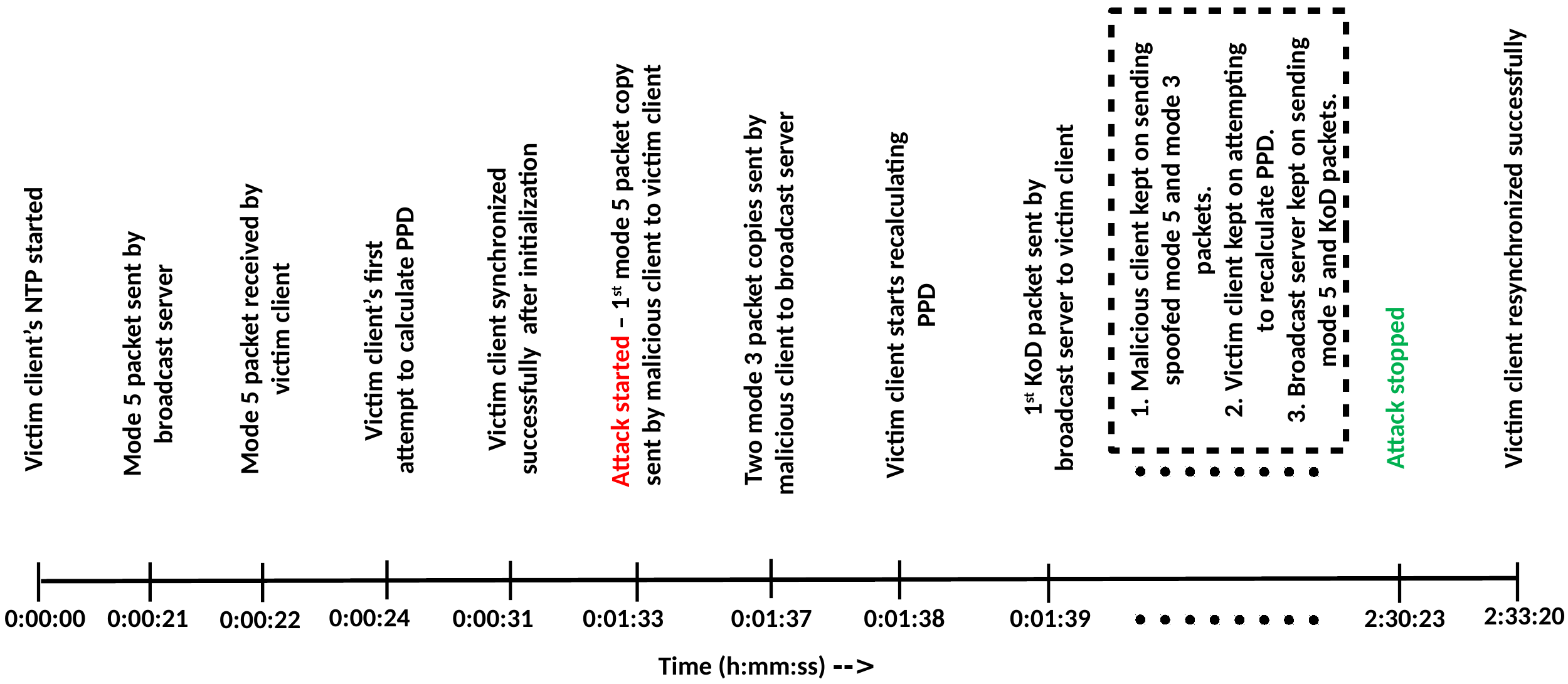}}
		\caption{Timing of various events. (PPD: Path Propagation Delay)}
		\label{timings}
	\end{center}
\end{figure*}

We executed the attack for approximately 2.5 hours on the testbed setup. Figure \ref{timings} shows the relative timing of various events as seen in the experiment. 
The broadcast server sent a mode 5 packet at \texttt{0:00:21} which victim client received at \texttt{0:00:22}. Subsequently, the victim client made its first attempt to calculate path propagation delay at \texttt{0:00:24}. After this, the victim client successfully synchronized its clock with the broadcast server at \texttt{0:00:31}. The malicious client started the attack execution at \texttt{0:01:33} by sending mode 5 packets to the victim client which were spoofed to look like they are coming from the broadcast server. At \texttt{0:01:37}, the malicious client sent a burst of two mode 3 packets to the broadcast server which were spoofed to look like they are coming from the victim client. This forced the broadcast server to trigger the rate-limiting mechanism and send a KoD packet to the victim client. Meanwhile, at \texttt{0:01:38}, the victim client also sent mode 3 packets to the broadcast server in order to recalculate the path propagation delay. At \texttt{0:01:39}, the broadcast server sent first KoD packet with polling interval of 64 seconds to the victim client due to the rate with which it was receiving mode 3 queries (from both victim client and malicious client). Between \texttt{0:01:39} and \texttt{02:30:23}, the malicious client kept on sending spoofed mode 3 and mode 5 packets to the broadcast server and victim client respectively in order to continue the attack while the broadcast server kept on sending genuine mode 5 packets and KoD packets to the victim client. Meanwhile, the victim client made 30 (unsuccessful) attempts\footnote{Victim client, on an average, sends four mode 3 packets to calculate path propagation delay in each attempt.} to calculate the path propagation delay in order to synchronize its clock with the broadcast server. At \texttt{02:30:23}, the malicious client stopped the attack execution and subsequently, after 177 seconds, the victim client could successfully resynchronize its clock with the broadcast server in its 31st attempt. During the attack execution, the number of different types of NTP packets sent by each entity is shown in Table \ref{each_entity_packet}.

\begin{table}[h]
	\centering
		\caption{NTP packets sent by each entity. (NA: Not Applicable)}
	\begin{tabular}{|c|c|c|c|}
		\hline
		\textbf{Entity} & \textbf{Mode 3} & \textbf{Mode 5} & \textbf{KoD} \\ \hline
		Malicious Client & 887 & 8388 & NA \\ \hline
		Victim Client & 125 & NA & NA \\ \hline
		Broadcast Server & NA & 138 & 886 \\ \hline
	\end{tabular}
	\label{each_entity_packet}
\end{table}

\subsection{Effectiveness of Attack in Authenticated Multicast Mode}
RFC 5905 \cite{rfc5905} allows NTP clients and server to operate in multicast mode as well. The multicast mode operation is similar to that of the broadcast mode. Our proposed attack can also prevent a client configured in multicast mode from synchronizing itself with a multicast NTP server. To test the attack in multicast NTP mode, we configured victim client to listen mode 5 packets by adding \texttt{multicastclient} keyword and IPv4 multicast group address \texttt{224.0.1.1} in its NTP's configuration file. We also modified NTP broadcast server's configuration file by replacing the network broadcast address with \texttt{224.0.1.1} to send mode 5 NTP packets on this address. NTP daemon of broadcast server and victim client were then restarted to apply the changes. We then let the broadcast server and victim client synchronize to external stratum 2 server on the Internet and broadcast server respectively. To launch the attack, malicious client simply sent copies of mode 5 and mode 3 NTP packets with appropriate source and destination IP addresses as discussed earlier. On receiving mode 5 packets, victim client made several attempts to recalculate the path propagation delay, however, several mode 3 packets sent by the malicious client to broadcast server prevented the victim client to calculate the delay each time. Thus, the proposed attack was found to be effective in NTP's authenticated multicast mode also.

\subsection{Attack Surface}

To quantify the effect of proposed attack, it is important to know the attack surface space. Thus, we scanned entire IPv4 address space\footnote{excluding private and reserved IP addresses (covering approximately 13.8\% of the entire IPv4 address space).} to determine the usage of NTP's broadcast mode in the wild. We also measured the usage of different \textit{ntpd} versions on Internet. In this subsection, we discuss the scanning testbed setup, measurement procedure and the obtained results.

\subsubsection{Setup}

We performed the Internet wide scan using a Virtual Machine (VM) on Google Cloud Platform. This VM was running Ubuntu 18.04 operating system and having one virtual CPU with 6.5 GB of physical memory. This  VM also had \textit{Python 2.7.0} and \textit{Scapy} \cite{scapy_documentation} libraries to send NTP control packets in Internet using raw sockets. We also installed \textit{tcpdump} to capture the resulting NTP traffic at VM. The captured NTP traffic was then processed and parsed using \textit{Python} and \textit{Scapy} libraries to extract details such as NTP association type, \textit{ntpd} version, stratum level, etc.  

\subsubsection{Measurement Methodology}

We first executed \texttt{peers} command on a machine and captured the resulting NTP control packets. The \texttt{peers} command is used to obtain a list of all associations used by an NTP client. Once we had the required packets, we used a python script that creates raw sockets and sends the payload of captured NTP control packets to the entire IPv4 address space. We captured the resulting NTP traffic using \textit{tcpdump} in the form of a \texttt{pcap} file. This measurement was taken from September 20th to September 22nd, 2019 and we obtained the responses from 1,171,607 IP addresses. After this, we rescanned these responding IP addresses using \texttt{readvar} command. This command, when executed for a particular association, returns the information such as \textit{reference ID}, \textit{stratum}, \textit{reference timestamp}, \textit{mode of association}, etc.  This measurement was taken between September 23rd and September 24th, 2019. We obtained the responses from 722,217 IP addresses and we consider only these addresses here. 

\subsubsection{Results}

The results for the measurement of broadcast association usage in the wild are shown in Table \ref{assoc_count}.
\begin{table}[h]
	\centering
	    \caption{Usage of different NTP modes in Internet}
	\begin{tabular}{|c|c|}
		\hline
		\textbf{Detail} & \textbf{Count} \\ \hline
		Hosts with at least 1 broadcast association & 2740 \\ \hline
		Total broadcast associations & 8090 \\ \hline \hline
		Hosts with at least 1 symmetric association & 55712 \\ \hline
		Hosts with at least 1 client/server association & 668,097 \\ \hline
		Total symmetric associations & 130,249 \\ \hline
		Total client/server associations & 1,273,892 \\ \hline
	\end{tabular}
	\label{assoc_count}
\end{table}
On a side note, we also measured the usage of symmetric as well as client/server associations in the wild. The results for these measurements are also shown in Table \ref{assoc_count}.
\\\\
\textbf{Measurement of \textit{ntpd} versions being used in the Internet:} We also measured the usage of different \textit{ntpd} versions in the Internet. For this purpose, we rescanned 722,217 IP addresses that responded in the previous measurement using \texttt{readvar} command. This command, when executed for a system as a whole, returns information such as \textit{ntpd version}, \textit{operating system}, etc. This measurement was taken from September 25th to September 26th, 2019 and we obtained the responses from 648,187 IP addresses. Out of these 648,187 IP addresses, only 496,079 IP addresses returned responses having version information. The results for this measurement is shown in Table \ref{version_count}.
\begin{table*}[h]
	\centering
	\caption{\textit{ntpd} versions being used in Internet. \textit{Others}: Either other older versions or unreadable version number}
	\begin{tabular}{|c|c|c|c|c|c|c|c|c|c|}
		\hline
		\textbf{Version} & 4.2.0 & 4.2.8 & 4.2.6 & 4.1.1 & 4.2.4 & 4.2.7 & Just ``ntpd v4" & Others & Total \\ \hline 
		\textbf{Count} & 18024 & 13476 & 8166 & 7308 & 1203 & 569 & 446620 & 713 & 496,079 \\ \hline 
	\end{tabular}
	\label{version_count}
\end{table*}
We can notice from the table that the most recent \textit{ntpd} version 4.2.8 is the second most popular (after 4.2.0) in the Internet. Since we found the vulnerability in four latest \textit{ntpd} 4.2.8 subversions - 4.2.8p10, 4.2.8p11, 4.2.8p12 and 4.2.8p13, we also measured how many systems on the Internet are using these subversions. These results are shown in Table \ref{latest_version_count} in descending order.
\begin{table}[h]
	\centering
	\caption{Recent subversions of \textit{ntpd} 4.2.8 being used in Internet. \textit{Others}: Either other subversions or just \textit{ntpd} 4.2.8}
	\begin{tabular}{|c|c|}
		\hline
		\textbf{Version} & \textbf{Count} \\ \hline
		4.2.8p10 & 2881 \\ \hline
		4.2.8p12 & 2333 \\ \hline
		4.2.8p11 & 1960 \\ \hline
		4.2.8p13 (latest) & 1756 \\ \hline
		Others & 4546 \\ \hline
		Total & 13476 \\ \hline
	\end{tabular}
	\label{latest_version_count}
\end{table}
\\\\
\textbf{Measurement of \textit{ntpd} versions being used by broadcast clients in the Internet:} We also measured which different \textit{ntpd} versions are being used by broadcast clients in the Internet. Out of 2740 IP addresses using at least one broadcast association, we obtained response containing version information from 1552 IP addresses only. The results for this measurement are shown in Table \ref{version_count_broadcast}. Similarly, the results for the usage of \textit{ntpd} subversions 4.2.8p10, 4.2.8p11, 4.2.8p12 and 4.2.8p13 by NTP broadcast clients in the Internet are shown in Table \ref{latest_version_count_broadcast}.
\begin{table}[h]
	\centering
	\caption{\textit{ntpd} versions being used by systems with at least 1 broadcast association in Internet. \textit{Others}: Either other older versions or unreadable version number}
	\begin{tabular}{|p{1cm}|c|c|c|c|c|c|c|}
		\hline
		\textbf{Version} & 4.2.8 & 4.1.0 & 4.2.0 & 4.1.1 & 4.2.6 & Just ``ntpd v4" & Total \\ \hline
		\textbf{Count} & 23 & 4 & 3 & 2 & 1 & 1519 & 1552 \\ \hline
	\end{tabular}
	\label{version_count_broadcast}
\end{table}
\begin{table}[h]
	\centering
	\caption{Recent subversions of \textit{ntpd} 4.2.8 being used by systems with at least 1 broadcast association in Internet. \textit{Others}: Either other subversions or just \textit{ntpd} 4.2.8}
	\begin{tabular}{|c|c|}
		\hline
		\textbf{Version} & \textbf{Count} \\ \hline
		4.2.8p10 & 10 \\ \hline
		4.2.8p13 (latest) & 3 \\ \hline
		4.2.8p12 & 5 \\ \hline
		4.2.8p11 & 1 \\ \hline
		Others & 4 \\ \hline
		Total & 23 \\ \hline
	\end{tabular}
	\label{latest_version_count_broadcast}
\end{table}

\subsection{How Significant is the Attack Surface?} 

We can notice from Table \ref{assoc_count} that the number of broadcast associations (i.e. 8090) is quite less as compared to client/server associations (i.e. 1,273,892) and thus, one can consider that the attack surface is insignificant. However, we argue that the actual attack surface depends not only on the number of broadcast associations but also on the stratum of NTP broadcast hosts on Internet. This is because if the victim NTP broadcast host is under attack and is not able to synchronize its clock with a broadcast server, eventually it will not be able to provide time information to other NTP hosts for whom it is acting as an NTP server. Thus, we performed a measurement study (using the output of same \texttt{readvar} command) to check the stratum of broadcast associations on the Internet and the number of NTP hosts synchronizing their clock time from these servers. Surprisingly, we found that 8088 out of 8090 broadcast associations are stratum 2 (considered as highly accurate) associations such that the number of stratum 3 NTP hosts taking time from these stratum 2 broadcast clients is 433. Subsequently, these stratum 3 NTP hosts provide time to stratum 4 NTP hosts and so on. Thus, if the lower stratum NTP hosts are attacked, it will directly affect the whole NTP hierarchy which makes the attack surface significant.

We would also like to point that most of the NTP clients on the Internet do not respond to the \texttt{readvar} command that we used to measure the attack surface. This is because \texttt{readvar} command is used to launch popular NTP amplification/reflection attack \cite{czyz2014taming}. Due to this shady image, network administrators usually disable this remote query or deploy firewalls to drop it \cite{malhotra}. This is one of the primary reason why only 722,217 out of 1,171,607 NTP clients (i.e., only 61.64 \%) on Internet responded back to \texttt{readvar} command. Also, we showed that only 23 out of 2740 broadcast NTP clients (i.e., 1.20 \%) in Internet responded back with exact version information. Remaining 1519 broadcast NTP clients returned back response with version as ``ntpd v4" (superset for \textit{ntpd} v4.*.*p*) only. 

Moreover, NTP's broadcast mode is typically implemented when the broadcast server and clients are present behind Network Address Translation (NAT) boxes which prevents them from being scanned from external machines like ours. \textbf{Thus the exact number of  NTP broadcast clients in the wild could not be measured and the presented results show only the floor value of the overall attack surface}. Also, the proposed attack is effective against latest \textit{ntpd} versions and sooner or later, several NTP clients in Internet will update their \textit{ntpd} implementations which eventually makes them vulnerable to the attack.  

\section{Possible Countermeasures}
\label{countermeasures}

In the last two sections, we showed that our proposed attack can prevent an NTP client from synchronizing its clock with an NTP broadcast server and that the overall attack surface in the Internet can not be neglected. Thus, we also suggest possible mitigation approaches that can be deployed to counter the proposed attack until the vulnerable protocol implementations are appropriately patched. These countermeasures are as follows:
\\\\
\textbf{Out-of-band Path Propagation Delay Calculation:}  Our proposed attack requires malicious client to send several mode 3 NTP queries to broadcast server which are spoofed to look like they are coming from a genuine NTP client. These mode 3 queries cause broadcast server to send KoD packets instead of mode 4 responses to the genuine client. This results into genuine client's failure to calculate path propagation delay. For this reason, we suggest that NTP broadcast clients should use an out-of-band path propagation delay calculation technique. For instance, broadcast clients can exchange a couple of TCP or UDP packets with the broadcast server at regular intervals to calculate mean round-trip time. However, this recommendation requires making changes to RFC 5905. Moreover, this approach leads to higher bandwidth consumption at both broadcast server and client side. 
\\\\
\textbf{Relying Exclusively on Broadcast Mode is Harmful:} The easiest way to prevent proposed attack is not to exclusively rely on broadcast mode for synchronization purpose. Instead, it should be made sure that the clients are backed by one or more NTP servers using other modes of NTP operation such as client/server mode or symmetric mode. In this way, the NTP clients have the option of choosing other existing associations when the broadcast association fails.
\\\\
\textbf{Prevent IP Spoofing:} Since the proposed attack requires malicious client to use spoofed IP addresses to send mode 5 and mode 3 packets to victim client and broadcast server respectively, any IP spoofing mitigation approach can prevent the proposed attack. Thus, techniques such as ingress/egress filtering \cite{rfc2827, Wang2007, Jin2003, hubballi2017event} can be implemented on the border routers to block malicious traffic. However, this approach will fail to mitigate the attack if it is launched within a local network.
\\\\
\textbf{Limiting Access of NTP Broadcast Clients:} RFC 5905 suggests ``to limit the access of NTP clients to known or trusted NTP broadcast servers as it will prevent malicious traffic from reaching the NTP clients" \cite{rfc5905}. Due to this, the malicious client will not be able to send spoofed mode 5 packets to the victim client which will subsequently prevent the attack.

\section{Prior Work}
\label{ntp_literature}

In this section, we describe earlier works wherein authors presented attacks against clock synchronization using NTP. Subsequently, we also discuss known defense approaches to counter these attacks.

\subsection{Attacks against NTP}

Malhotra and Goldberg \cite{malhotra} proposed two attacks against NTP's broadcast mode. The first attack is a replay attack wherein a Man-in-the-Middle (MitM) malicious client between an NTP broadcast server and a victim client replays a contiguous sequence of broadcast packets at regular interval to the victim client. This causes the victim's clock to stuck at a particular time. The second attack is a DoS attack wherein a malicious client can cause an error in the NTP operation of victim client by sending a badly authenticated mode 5 packet. As a result, the victim client breaks its association with the broadcast server. Following attack disclosure, appropriate patches have been added to \textit{ntpd} v4.2.8p6 and later versions to mitigate these attacks \cite{malhotra}.

In another work \cite{malhotra_ndss}, authors proposed several attacks against NTP's unauthenticated client/server mode. To launch the first attack \cite{malhotra_ndss}, a malicious client sends bogus small time shifts to a victim client and then, when the malicious client is ready, it sends a big time shift to the victim client to create DoS. Following attack disclosure, vulnerable \textit{ntpd} versions were patched to mitigate this attack \cite{malhotra_ndss}. To launch the second variant, an off-path malicious client sends a spoofed KoD packet to the victim NTP client for each of the client's pre-configured NTP servers. On receiving KoD packets, the client immediately stops sending NTP packets to the servers due to which it is not able to synchronize its local clock. After public disclosure, newer \textit{ntpd} versions are made resilient to this attack by patching the origin timestamp validation process. In one of the attacks, authors also showed how certain IPv4 fragmentation policies used by a client's and server's operating systems can be exploited to hijack an unauthenticated NTP connection established between the client and server.

In \cite{aanchal}, authors discussed an attack wherein a malicious client obtains the transmit timestamp $T_3$ of a pending mode 3 query sent by a client to a NTP server and then crafts a fake mode 4 response having origin timestamp as $T_3$ and bogus time information. Since this response is having valid origin timestamp, the client accepts and processes the response due to which its clock is shifted to a wrong time. This vulnerability is not yet patched \cite{aanchal} and thus, the current versions of \textit{ntpd} are still vulnerable. One more vulnerability was discussed in this work wherein a malicious client can bypass TEST2 by spoofing server's mode 4 response packets with their origin timestamp set to zero. Following attack disclosure, this vulnerability has been fixed in \textit{ntpd} v4.2.8p9 and later versions \cite{aanchal}.

\subsection{Defense Approaches}

\textbf{Cryptographic Techniques:} Some approaches \cite{aanchal, 197241} use cryptographic techniques to authenticate NTP packets. However, NTP traffic is very rarely authenticated in practice \cite{malhotra_ndss} because of various reasons such as cumbersome key distribution mechanism, weaknesses in the Autokey protocol \cite{rfc5906} for public-key authentication, etc. \cite{rfc7384}. This leads to the development of NTPsec \cite{ntpsec1, ntpsec2}. Unfortunately, the adoption of NTPsec is still very slow and moreover, authentication and encryption do not mitigate MitM attacks as an MitM adversary can simply drop traffic destined to port 123 (default for NTP).
\\\\
\textbf{Path Redundancy:} A class of work in the literature utilize path redundancy on the Internet to avoid MitM adversaries \cite{mizrahi2012slave, 6644754}. Under this approach, multiple paths on the Internet are used to connect NTP clients and servers. Thus, even if one of the paths between an NTP client and a server is compromised, the client is able to synchronize its clock by exchanging NTP packets over other paths. The drawback of this approach is that it can not mitigate attacks which do not require a MitM position (our proposed attack, for example).
\\\\
\textbf{Server Redundancy:} Deutsch et al. \cite{deutsch2018preventing} proposed a new NTP client \textit{Chronos} which first generates server redundancy by creating large server pools and then carefully samples servers in these pools. Since \textit{Chronos} synchronizes its clock with the help of large server pools, even a malicious client with MitM position can not stop the client from obtaining correct time information from other servers. As a result, this approach is resilient to \textit{on-path attacks}. In case of \textit{off-path attacks}, \textit{Chronos'} clock selection algorithm rejects wrong time information received from a malicious adversary as it differs from the clock values of servers present in the pool. Thus, \textit{Chronos} is resilient to \textit{off-path} attacks as well. However, this setup requires changes in the core NTP infrastructure of the Internet to counter the attacks.

\section{Conclusion}
\label{ntp_conclusion}

In this paper, we proposed a new attack that can prevent a client from synchronizing its clock to a broadcast NTP server. We tested the proposed attack in real network and reported the results. The proposed attack was found to be effective in both authenticated and unauthenticated broadcast and multicast modes. We also discussed how the proposed attack can render all the broadcast clients in the network unsynchronized. We measured the attack surface by scanning the entire IPv4 address space using extensive experiments and showed that several highly accurate low stratum hosts on Internet are vulnerable to the attack. We also suggested few countermeasures that can be deployed as a first line of defense to mitigate the proposed attack. Moreover, we are also working on an effective prevention approach to counter this attack. We believe that our new attack disclosure will motivate researchers in the security community to closely evaluate the protocol specification further, identify new vulnerabilities and propose appropriate patches or defense mechanisms to make the protocol and its implementations secure and robust. 

\bibliographystyle{IEEEtran}
\bibliography{sample-base}

\end{document}